\documentclass[fleqn,twoside]{article}
\usepackage{espcrc2}

\title{MONOPOLE FIELDS FROM VORTEX SHEETS RECONCILING ABELIAN AND
CENTER DOMINANCE}

\author{J. Fr\"{o}hlich\address{Theoretical Physics, ETH-
H\"{o}nggerberg ,
CH-89093, Z\"{u}rich, Switzerland} and P.A.
Marchetti\address{Dipartimento di
Fisica, Universit\`a di Padova and INFN-Sezione di Padova, I-35131
Padova,
Italy}\thanks{This work was supported by the European Commission RTN
programme HPRN-CT2000-00131. Talk presented by P.A. Marchetti}}

\begin{document}

\begin{abstract}

We describe a new order parameter for the confinement-deconfinement
transition
in lattice SU(2) Yang-Mills theory. It is expressed in terms of magnetic
monopole field correlators represented as sums over sheets of center
vortices. Our construction establishes a
link between ``abelian" and ``center dominance".
It avoids an inconsistency in the treatment of small scales present in
earlier definitions of monopole fields by respecting Dirac's
quantization condition
for magnetic fluxes.
\vspace{1pc}
\end{abstract}

\maketitle
\section{Abelian and center dominance}
It is widely believed that confinement in SU(2) Yang-Mills theory is due
to
condensation of topological defects, but it is still debated whether the

relevant defects are center vortices or magnetic monopoles.

In the first scenario \cite{vor} confinement is usually discussed in
terms of
area decay for the Wilson loop. In lattice theory, the location of
vortex sheets
giving rise to area law have been identified with surfaces of
plaquettes, $p$, where $ sign [Tr U_{\partial p}] = -1$ (thin vortices),
\cite {thin}.
Here $U_\mu$ denotes the $SU(2)$-gauge field, and $U_{\partial p}$ the
Wilson
plaquette, [more precisely, if $p$ is described by a lattice site $x$
and two
direction $\mu,\nu$ then $
U_{\partial p} \equiv U_{\mu\nu} (x) \equiv U_\mu (x) U_\nu (x+ \hat
\nu)
U^\dagger (x + \hat\nu) U^\dagger_\nu (x)$].
However it was later proved \cite{mack} that close to the continuum, at
zero
temperature,  thin vortices form a dilute gas and hence they are unable
to
induce  an area decay of the Wilson loop.
Therefore, to explain confinement, one needs to invoke either ``thick
vortices" \cite{fat}, where  $sign [Tr U_L] =-1$ for loops $L$
comprising more than four links, or, presumably equivalently
\cite{equiv}, $P$-vortices which are defined
as follows \cite{pvor}: one introduces a gauge fixing (maximal center
gauge)
by maximizing
$\sum_{x,\mu} (Tr U_\mu (x))^2.$
In this gauge one defines the ${\bf Z}_2$-gauge field $Z_\mu (x) = sign[
Tr
U_\mu (x)]$; the location of $P$-vortices is identified with the set of
plaquettes where
$Z_{\partial p}= Z_{\mu\nu} (x) =-1$.
We refer to the above scenario, where the center degrees of freedom are
believed to be the relevant ones for confinement,
as ``center dominance".

't Hooft put on a concrete basis a proposal explaining confinement as a
condensation of magnetic monopoles \cite{mon} : he suggested \cite{'th}
to
construct a scalar field $X(U)$ with values in $su(2)$, as a function
of the gauge field $U$ and transforming in the adjoint representation of
the
gauge  group $SU(2)$. By requiring that $X(U)$ be diagonal he then fixes
a gauge (``abelian projection"). The resulting theory exhibits  a
residual $U(1)$
gauge invariance.

The argument of the diagonal component of the $SU(2)$ gauge field in
this
``abelian projection gauge" plays the role of a compact $U(1)$ ``photon"
field,
$A_\mu$, with range ($-2\pi, 2\pi$), and the off-diagonal components are

described by a complex field, $c$, charged with respect to the residual
$U(1)$ gauge group. The points in space-time where the two eigenvalues
of the
matrix $X$ coincide identify the positions of the monopoles in this
gauge.
Confinement is believed to emerge as a consequence of
monopole--condensation in the form of a ``dual Meissner effect''.

We refer to this scenario for confinement (together with the assumption
that
off-diagonal degrees of freedom are irrelevant for a description of low
energy physics) as ``abelian dominance".

\section{U(1) monopole order parameter}
In \cite{mord}, a monopole field operator $\hat m$ is proposed which
plays the role  of an order parameter for the confinement--deconfinement

transition in the ``abelian dominance" scenario, i.e., with vanishing
v.e.v.
in the deconfinement phase and v.e.v. $\langle \hat m \rangle \not = 0$
in
the confinement phase.

A euclidean representation of the two-point monopole correlation
function
$G(x,x^\prime) =\langle \hat m (x) \hat m^\dagger (x^\prime) \rangle$ is

constructed as follows: let $E^x_\mu, x = (x^0, \vec x)$ denote the
(lattice)
electric Coulomb field generated on the 3--dimensional sublattice,
${\bf Z}^3_{x^0}$, at constant euclidean time $x^0 \in {\bf Z}$, by a
unit
charge located at $\vec x \in {\bf Z}^3$. Denoting by $\Delta_{x^0}$ the

3--dimensional (lattice) laplacian in ${\bf Z}^3_{x^0}$, one has:

\begin{equation}
E_i^x (y) = \partial_i \Delta^{-1}_3 (y, x), i= 1,2,3 \\,  \\ E^x_0 (y)
=0,
\end{equation}
for $y \in {\bf Z}^4$ with $y^0=x^0$, and $E_\mu^x(y)=0$ elsewhere, so
that $\sum_\mu \partial_\mu E^x_\mu(y)=\delta_x(y)$.
Let $B^x_{\nu\rho\sigma}$ denote the * dual of $E^x_\mu$, supported on
cubes
in the dual lattice, let $j_\mu^{xx^\prime}$ be a unit current supported
on a
path from $x$ to $x^\prime$ [$\sum_\mu \partial_{\mu} j_\mu^{xx^\prime}
(y)
= \delta_x (y) - \delta_{x^\prime} (y) $], and let
$\omega^{xx^\prime}_{\nu\rho\sigma}$ denote its * dual, supported on
cubes dual to the links in the support of $j^{xx^\prime}_\mu$.

The Yang-Mills action is defined by

\begin{equation}
\label{ym}
S (U) = - \beta \sum_p Tr U_{\partial p} = - \beta \sum_{y,\mu,\nu}
Tr U_{\mu\nu} (y),
\end{equation}
and we define a modified action $S_{\omega^{xx^\prime}} (U| X, B^x,
B^{x^\prime})$, depending on a unit-norm $su (2)$ scalar $X$ determining
an
abelian projection, by multiplying $U_{\mu\nu}(y)$ in the plaquette term
in (\ref{ym}) by

\begin{equation}
\label{pla}
 e^{i X(y) 2\pi \sum_{z,\rho} \partial_\rho \Delta^{-1} (y,z)
[\omega_{\mu\nu\rho}^{xx^\prime}
 + B^x_{\mu\nu\rho}-B^{x^\prime}_{\mu\nu\rho}](z)}
\end{equation}
where $\Delta$ is the 4-d lattice laplacian.

The 2-point monopole correlation function corresponding to $X$ proposed
in
\cite{mord} can be defined as the Yang-Mills v.e.v. of the disorder
field

\begin{equation}
\label{dis}
D_{\omega^{ xx^\prime}} (B^x, B^{x^\prime}) = e^{- [S_{\omega^{
xx^\prime}}
(U| X, B^x, B^{x^\prime}) - S(U)] }.
\end{equation}
Since $X$ transforms under the adjoint representation, the v.e.v. of $D$
is
$SU(2)$--gauge invariant.

To make definition (\ref{dis}) plausible, we notice  that, in an abelian
projection
gauge, and after integrating out the ``charged field'' $c$, the $SU(2)$
theory appears
as a $U(1)$-gauge theory. In this $U(1)$ theory, the disorder field
(\ref{dis}) is constructed \cite{pol} by translating the field strenghth
of
$A_\mu(y)$ by $\pi \sum_{z,\rho} \partial_\rho \Delta^{-1} (y,z)
[\omega^{xx^\prime}  + B^x - B^{x^\prime}  ]_{\mu\nu\rho}(z)$ in the
action.
Wegner-'t Hooft duality \cite{dual} maps a pure $U(1)$ gauge theory onto

a non--compact abelian Higgs (n.c.H.) model, exchanging the role of
monopoles
and charges. One can prove \cite{fm} that for a $U(1)$ version of the
disorder field (\ref{dis}), one has the duality:

\begin{eqnarray}
\label{du}
\nonumber
&\langle D_{\omega^{ xx^\prime}} (B^x, B^{x^\prime})\rangle_{U(1)} =\\
& \langle e^{i (\theta (x) - \theta (x^\prime))}e^{i\sum_{y,\mu}(E^x_\mu
- E_\mu^{x^\prime}) (y)A_\mu (y)}\rangle_{n.c.H.}
\end{eqnarray}
where $\theta$ is the charged field and $A_\mu$ the gauge field of the
dual Higgs model.
The r.h.s. of (\ref{du}) appears as the two-point correlation function
of the non-local gauge-invariant charged field of the Higgs model
constructed according to Dirac's ansatz \cite{di}.
If we expand the l.h.s. [r.h.s.] of (\ref{du}) in terms of worldlines of
monopoles [charges], such worldlines have a source at $x$ and a sink at
$x^\prime$. The $B [E]$ current distributions emerging from these points
describe a cloud of soft ``photons".
Notice that choosing $\omega_{\mu\nu\rho}^{xx^\prime} =
\omega^x_{\mu\nu\rho}- \omega^{x^\prime}_{\mu\nu\rho}$, with $\omega^x$
dual to a path at constant time $x^0$ from $x$ to $\infty$ ( where
suitable b.c. are imposed so that still $\partial_\mu (*
\omega^{xx^\prime})_\mu=\delta_x- \delta_{x^\prime}$) one recognizes the
sum $B^x_{\mu\nu\rho} + \omega^x_{\mu\nu\rho}$ as the lattice magnetic
field at euclidean time $x^0$ of a monopole located at $\vec x$, with
$\omega^x_{\mu\nu\rho}$ playing the role of its Dirac string.
It has been rigorously proved in \cite{fm} that as $|x-x^\prime|
\rightarrow \infty$, the correlation function (\ref{du}) tends to 0 in
the deconfined phase of the U(1) gauge theory [Coulomb phase of the dual
Higgs model] and approaches a finite value in the confined phase
[superconducting Higgs phase]. Hence the monopole field operator
reconstructed from the v.e.v. of the disorder field correlation
(\ref{du}) in U(1) theory is a good order parameter for the
confinement-deconfinement transition.
On the basis of these arguments it has been claimed in \cite{mord} that
the monopole field operator reconstructed from v.e.v. of the disorder
field  (\ref{dis}) is a good order parameter in SU(2) Yang-Mills theory.
Numerical evidence in favour of this conjecture emerged in \cite{dig};
(see also \cite{pol} \cite{poli}). Critical exponents associated with
this transition extracted from the behaviour of the v.e.v. of
(\ref{dis}) appear to be independent of the choice of $X$ \cite{dig1}.

\section{Inconsistency and cure}

In spite of its great numerical success, the order parameter based on
(\ref{dis}) is inconsistent in the treatment of small scales, because it
violates
Dirac's quantization of fluxes required  for self-consistency of a
theory where dynamical charges (in our case represented by $c$) and
monopoles coexist. This inconsistency shows up in an unphysical
dependence on the ``Dirac string'' $\omega^{xx^\prime}$ exhibited by
$\langle D_{\omega^{xx^\prime}} (B^x, B^{x^\prime})\rangle$. In the
abelian projection gauge-fixed theory, this feature appears because the
U(1)-gauge theory obtained by integrating out $c$ has a dual which is a
compact Higgs model, with dynamical charges \underline{and} monopoles.
The 2-point correlation function of the charged field constructed
according to Dirac's ansatz then depends on the choice of the Dirac
surfaces swept out by Dirac strings attached monopole worldlines.
Let us explain how this happens \cite{fm1}. In the compact dual Higgs
model, the Dirac surfaces, $S$, are described by integer-valued  surface
currents, $n_{\mu\nu}$, supported on the plaquettes dual to $S$. A
change of Dirac surfaces, $S\rightarrow S^\prime$, for a fixed
configuration of monopole worldlines, corresponds to the shift

\begin{equation}
\label{shi}
n_{\mu\rho} \rightarrow n_{\mu\rho}+ \partial_\mu V_\rho - \partial_\rho
V_\mu,
\end{equation}
where $V_\mu$ is the integer current supported on the dual of the cubes
contained in the volume whose boundary is the closed surface $S^\prime -
S$.
In the partition function, the interaction of the electric currents
generated by the charged particles, whose worldlines are described by an
integer 1-current $j_\mu$, with the Dirac surfaces of the monopoles is
of the form

\begin{equation}
\label{int}
ieg \sum_{y,z,\mu,\rho} j_\mu (y) \partial_\rho \Delta^{-1} (y,z)
n_{\rho\mu} (z)
\end{equation}
where $e$ is the electric charge of the matter field and $g$ the
magnetic charge of the monopole field.
The change (\ref{shi}) induces a shift of (\ref{int}) by

\begin{equation}
i e g \sum_{y,\mu} j_\mu (y) V_\mu (y)
\end{equation}
which when exponentiated is unity, as required, provided it is an
integer multiple of $2\pi i$ [Dirac quantization condition for fluxes].
This happens in the partition function if Dirac's quantization condition
for charges holds, i.e.  $e g = 2\pi q, q$ an integer, because $j_\mu$
and $V_\mu$ are integer currents. In the Dirac ansatz for the 2-point
function of the charged field, however, $j_\mu$ acquires additional
Coulomb-like terms, $E_\mu$, which are real-valued, hence

\begin{equation}
eg \sum_{y,\mu} E_\mu (y) V_\mu (y) \notin 2\pi {\bf Z}
\end{equation}
even if $e g \in 2\pi {\bf Z}$, and the Dirac strings of monopoles
become unphysically ``visible".
An obvious cure for this inconsistency  would be to replace the Coulomb
field $E^x_\mu$ by a ``Mandelstam string" $j_\mu^x$ \cite{man},
squeezing the entire flux of $E^x$ into a single line from $x$ to
$\infty$ at fixed  time (and adding suitable b.c.).

However, this squeezing of the flux is so strong that it produces IR
divergences [$(\sum_{y,z,\mu} (E^x_\mu - E_\mu^{x^\prime})(y)
\Delta^{-1} (y-z) (E^x_\mu - E^{x^\prime}_\mu) (z) < \infty$ but
$\sum_{y,z,\mu} (j^x_\mu - j^{x^\prime}_\mu) (y) \Delta^{-1} (y-z)
(j_\mu^x - j_\mu^{x^\prime}) (z) = \infty $].

To avoid these divergences, we need to replace a fixed Mandelstam string
by a sum over fluctuating Mandelstam strings weighted by a measure
${\cal D} \nu_q (j_\mu^x)$ such that, in the scaling limit,
\begin{eqnarray}
\label{cure}
\nonumber
\int {\cal D}\nu_q (j^x_\mu)& e^{i e\sum_{y,\mu}j_\mu^x (y) A_\mu
(y)}\sim\\
& e^{i e \sum_{y,\mu}
E_\mu^x (y) A_\mu(y)}
\end{eqnarray}
[The integer $q$ in the measure ${\cal D} \nu_q$ is the one appearing in
the Dirac quantization condition $eg= 2\pi q$].  It has been shown in
\cite{fm1}
that a measure with such properties can be constructed as follows:
Consider a 3-dimensional XY model supported on a lattice at constant
time $x^0$,
with the $U(1)$ spin field, $\chi$, of period $2 \pi q$ minimally
coupled, with charge $e$, to the compact gauge field $A_\mu$ of the
compact
Higgs model. Denote by $\langle \cdot \rangle^{x^0} (A)$ the
corresponding expectation value, with a coupling constant of the XY
model chosen
sufficiently large that the symmetry $\chi \rightarrow \chi + const$ is
spontaneously broken. The correlation functions of the field $\chi$ can
be
expressed in terms of ${\bf Z}/q-$ valued currents; in particular

\begin{eqnarray}
\label{ren}
\nonumber
&\big( \langle e^{i \chi(x)} e^{- i \chi(\infty)} \rangle^{x^0} (A)
\big)_{ren} \sim\\
& \int {\cal D} \nu_q (j^x_\mu) e^{i e \sum_{y \mu} j^x_\mu (y) A_\mu
(y)}
\end{eqnarray}
where $( \cdot )_{ren}$ involves a multiplicative renormalization taking
care of the selfenergies of Mandelstam strings.
${\cal D} \nu_q (j^x_\mu)$ is the measure with the desired properties.
This measure is supported on currents $ j^x_\mu$ associated with $q$
paths in a 3-plane at a fixed time starting at the site $x$ and reaching
infinity (``$\infty$''). Comparing (\ref{ren}) and (\ref{cure}) we see
that the
measure  ${\cal D} \nu_q (j^x_\mu)$ is peaked at $E^x_\mu$ at large
scales. The 2-point correlation function for the gauge-invariant charged
field in the compact abelian Higgs (c.H.) model is then given by

\begin{eqnarray}
\label{ch}
\nonumber
&\int {\cal D}\nu_q (j^x_\mu) \int {\cal D}\nu_q (j^{x^\prime}_\mu)
\langle e^{i (\theta (x) - \theta (x^\prime))}\\
&e^{i e \sum_{y,\mu}(j^x_\mu -
j_\mu^{x^\prime}) (y)A_\mu (y)} \rangle_{c.H.}
\end{eqnarray}
replacing the r.h.s. of (\ref{du}). This definition respects Dirac's
quantization condition for fluxes and, as a consequence, it is
independent of the Dirac strings of the magnetic monopoles of the
compact Higgs model. [See \cite{chord} for preliminary numerical
evidence for the validity of an order parameter for the Coulomb-Higgs
transition in this model, based on the above correlation function.]

The 2-point monopole correlation function obtained by duality from
(\ref{ch}) is given by
\begin{equation}
\label{mord1}
\int {\cal D}\nu_q (j^x_\mu) \int {\cal D}\nu_q (j^{x^\prime}_\mu)
\langle D(\Sigma( j^x-j^{x'}+j^{xx'})) \rangle
\end{equation}
and plays the role of the l.h.s. of (\ref{du}). Here $D(\Sigma)$ is the
't Hooft loop in the dual of the compact Higgs model corresponding to a
surface $\Sigma$ whose boundary is given by the support of $
j^x-j^{x'}+j^{xx'}$, with b.c. turning it into a closed curve.
$D(\Sigma)$ is obtained by shifting the field strength of $A_\mu$ by $2
\pi q * \Sigma_{\mu\nu}$ in the action, where $q \Sigma_{\mu\nu}$ is
the ${\bf Z}$-valued surface current supported on $\Sigma$. Since
$j^x_\mu $
is supported on $q$ paths, $\Sigma$ is a $q$-sheet surface with the $q$
sheets having a common boundary given by the single line support of
$j^{xx'}_\mu $.

\section{A new order parameter}

To export these ideas to $SU(2)$ Yang-Mills theory, one first remarks
that, in an abelian projection gauge, there appear a charged field, $c$,
of
electric charge 1 and monopoles of two species:
i) ${\bf Z}_2$-singular monopoles with magnetic charge \cite{deg} $g=2
\pi$, whose worldlines are defined
independently of the abelian projection. However they are screened
\cite{re} and dilute  close to the continuum at $T=0$ \cite{mack} and
thus cannot induce confinement;
ii) regular monopoles with magnetic charge $g=4 \pi$, whose worldlines
are only defined within the abelian projection gauge. It is the
condensation
of these monopoles that should be responsible for confinement, and for
them Dirac's quantization condition for charges is satisfied with $q=2$.

Therefore, we propose \cite{fm2} to construct the 2-point function for
such regular monopoles, $G^{YM} (x,x')$, as in equation (\ref{mord1})
for $q=2$, with the
following reinterpretation of notations: $\langle \cdot \rangle$ denotes
the expectation value in $SU(2)$ Yang-Mills theory and $D(\Sigma)$ is
the
$SU(2)$-'t Hooft loop which is defined by replacing the plaquette term
in (\ref{ym}) by
\begin{equation}
\label{dis1}
Tr \Bigl(U_{\mu\nu} (y) e^{i \sigma_3 2\pi *\Sigma_{\mu\nu}(y)}\Bigr).
\end{equation}

$G^{YM} (x,x')$ is thus defined as a sum of 't Hooft loops, and the
surfaces $\Sigma$ involved have 2 connected boundaries, each at constant
time, with fixed points the location of creation and annihilation, $x$
and $x'$, of the monopole. The definition
of $G^{YM} (x,x')$ is clearly intrinsic to $SU(2)$ Yang-Mills theory,
independent of the choice of an abelian projection. In an abelian
projection gauge, however, the surfaces $\Sigma$ are viewed as 2-sheet
surfaces of center vortices, with the two sheets joining along the
support of $j^{xx'}$, which becomes the worldline of a regular monopole.
Hence, whereas the definition of the worldline of a regular monopole
necessitates the introduction of an abelian projection, the positions of
creation and annihilation of the monopole are independent of it. There
is no semiclassical analogue of such monopoles in the $SU(2)$ theory
without abelian gauge fixing.

From correlation functions of regular monopole fields obtained
generalizing in obvious way the above definition, one can reconstruct a
monopole field
operator $\hat M$. We claim that its v.e.v. is a good order parameter
for the confinement-deconfinement transition. An argument supporting
this conjecture goes as follows: Since $2\Sigma_{\mu\nu}$ is
integer-valued, one can substitute $\sigma_3$ in (\ref{dis1}) with any
$su(2)$-valued field
$X$ of unit norm, selecting an abelian projection. Since the measure $
{\cal D}\nu_2 (j^x_\mu)$ is peaked near $E^x$, at large scales, one may
argue (using that $B^x=* E^x$) that, in the scaling limit, $G^{YM}
(x,x')$ behaves like the v.e.v. of the disorder operator (\ref{dis}) of
\cite{mord} , which, numerically, is a good order parameter. By
respecting the Dirac quantization condition for fluxes our construction
of $G^{YM} (x,x')$
avoids the inconsistency in the treatment of small scales of previous
monopole correlators and, although we expect that this inconsistency is
irrelevant for
the large distance behaviour controlling the phase transition, the
independence of $X$ of  $G^{YM} (x,x')$ could explain why, numerically,
the critical exponents of the transition have been found to be
independent of the choice of abelian projection \cite{dig1}.

Finally, our construction, being based on center vortex sheets, points
to a natural connection with the scenario of center dominance.
To make this more concrete, we replace the $SU(2)$-field $U_\mu$ by a
coset field $\bar U_\mu$, $SU(2)/{\bf Z}_2 \simeq SO(3)$-valued, and a
${\bf Z}_2 \simeq \{0,1\}$-valued 2-form $\sigma_{\mu\nu}$, obeying a
constraint \cite{mack} which admits a gauge-dependent solution:

\begin{equation}
\label{cons}
e^{i \pi \sigma_{\mu\nu}(y)} = sign[ Tr U_{\mu\nu} (y)] Z_{\mu\nu} (y)
\end{equation}
where $ Z_{\mu\nu} (y)$ is the Wilson plaquette of  $ Z_{\mu} (y)= sign
[Tr U_\mu (y)]$. The 't Hooft disorder field $D(\Sigma)$ is obtained in
terms of $\bar U_\mu$ and $\sigma_{\mu\nu}$ by shifting
$\sigma_{\mu\nu}$ by $* 2 \Sigma_{\mu\nu}$ in the action (in the
notation of (\ref{dis1})).
Plaquettes with a value $ -1$ for the first, gauge-invariant, term on
the r.h.s. of (\ref{cons}) identify the support of thin vortices; a
value $ -1$ for the second term in the center projection gauge
identifies the plaquettes in the support of $P$-vortices. There is
numerical evidence \cite{perc} that $P$-vortex sheets are percolating in
the space directions in the confinement phase, and this suggests that,
in the center projection gauge, the introduction of the vortex sheets
$\Sigma$, infinite in space directions, involved in the construction of
monopole correlation functions should be a small perurbation, and the $
{\cal D}\nu_2$ average of  $ \langle D(\Sigma) \rangle$ shoud not
vanish, whence  $\langle \hat M \rangle \neq 0$. In the deconfinement
phase at positive temperature, however, $P$-vortex sheets appear,
numerically, to be non-percolating in space directions \cite{perc}, and
one expects that the introduction of $\Sigma$ then leads to clustering,
implying   $\langle \hat M \rangle$=0. Condensation of regular monopoles
in the center projection gauge could then be interpreted as due to
percolation in space directions of $P$-vortex sheets. The relation
between worldlines of regular monopoles and vortex sheets, in our
construction, is a natural extension to open worldlines, with boundaries
corresponding to creation and annihilation of monopoles, of that
appearing in \cite{vormon} for closed monopole worldlines.

{\bf Acknowledgments.} Useful discussions with M. Chernodub, Ph. de
Forcrand, A. Di Giacomo, K. Langfeld and M. Polikarpov are gratefully
acknowledged.


\begin{thebibliography}{9}

\bibitem{vor} G. 't Hooft, Nucl. Phys. {\bf B153} (1979) 141;
G. Mack in ``Recent Developments in Gauge Theories", (Carg\'ese 1979),
't
Hooft ed. Plenum Press 1980;
A.M. Polyakov, Phys. Lett. {\bf 72B} (1978) 47 ;
H. Nielsen and P. Olesen, Nucl. Phys. {\bf B160} (1979) 360 ;
J.M. Cornwall, Nucl. Phys. {\bf B157} (1979) 392 .

\bibitem{thin} T. Yoneya, Nucl. Phys. {\bf B144} (1978) 195 .

\bibitem{mack}  G. Mack and V. Petkova, Z. Phys. {\bf C12} (1982) 177 .

\bibitem{fat} G. Mack and V. Petkova, Ann. Phys. {\bf 123} (1979) 442 ;
Ann. Phys. {\bf 125} (1979) 117 ; T.G. Kovacs and E.T. Tomboulis, Phys.
Rev. {\bf D57} (1998) 4504 .

\bibitem{equiv} M. Engelhardt and H. Reinhardt, Nucl. Phys. {\bf
B567}(2000) 249 .

\bibitem{pvor}  L. Del Debbio, M. Faber, J. Greensite and S. Oleynik,
Phys. Rev. {\bf D55} (1997) 2298 .

\bibitem{mon}  S. Mandelstam, Phys. Rep. {\bf 23C} (1976) 245; G 't
Hooft, in ``High Energy Physics'' EPS International Conference, Palermo
1975 ed. Zichichi.

\bibitem{'th}  G. 't Hooft, Nucl. Phys. {\bf B190} (1981) 455.

\bibitem{mord}  L. Del Debbio, A. Di Giacomo, G. Paffuti and P. Pieri,
Phys. Lett. {\bf B355},(1995) 255 ; G. Di Cecio, A. Di Giacomo,  G.
Paffuti and M.
Trigiante, Nucl. Phys. {\bf B489} (1997) 739 .

\bibitem{pol} L. Del Debbio, A. Di Giacomo and G. Paffuti, Phys. Lett.
{\bf B349} (1995) 513 ;M. Chernodub, M. Polikarpov, A. Veselov, Phys.
Lett. {\bf B399} (1997) 267 .

\bibitem{dual}  F. Wegner, J. Math. Phys. {\bf 12} (1971)  2254 ;
G. 't Hooft, Nucl. Phys. {\bf B138} (1978)  1 .

\bibitem{di}  P.A.M. Dirac, Can. J. Phys. {\bf 33} (1955) 650 .


\bibitem{fm}  J. Fr\"ohlich and P.A. Marchetti, Commun. Math. Phys.
{\bf 112} (1987) 343 .

\bibitem{dig}  A. Di Giacomo, B. Lucini, L. Montesi and G. Paffuti,
Nucl. Phys. Proc. Suppl. {\bf 63} (1998) 540 .

\bibitem{poli} M. Chernodub, M. Polikarpov and A. Veselov, Zh. Eksp.
Teor. Fiz. {\bf 96} (1989) 304 [JETP {\bf 69} (1989) 174].

\bibitem{dig1}  A. Di Giacomo, B. Lucini, L. Montesi and G. Paffuti,
Phys. Rev. {\bf D61} (2000) 034503, 034504 .

\bibitem{fm1}  J. Fr\"ohlich and P.A. Marchetti, Nucl. Phys. {\bf B511}
(1999) 770 .

\bibitem{man}  S. Mandelstam, Ann. Phys. {\bf 19} (1962) 1 .
\bibitem{chord} V.A. Belavin, M.N. Chernobub, F.V. Gubarev and M.I.
Polikarpov, talk presented at Lattice 2001-Berlin; V. Belavin, M.N.
Chernodub and M.I. Polikarpov hep-lat/0110150

\bibitem{deg} T. de Grand and T. Toussaint, Phys. Rev. {\bf D22} (1980)
2478; A.S. Kronfeld, G. Schierholz and U.J. Wiese, Nucl. Phys. {\bf 293}
(1987) 461 .

\bibitem{re} C. Hoelbling, C. Rebbi and V.A. Rubakov, Nucl. Phys. Proc.
Suppl. {\bf 83-84} (2000) 4853 .

\bibitem{fm2}  J. Fr\"ohlich and P.A. Marchetti, Phys. Rev. {\bf D64}
(2001) 014505 .

\bibitem{perc} M. Engelhardt, K. Langfeld, H. Reinhardt and O. Tennert,
Phys. Rev. {\bf D61} (2000) 054504 .

\bibitem{vormon}  L. Del Debbio, M. Faber, J. Greensite and S. Oleynik,
in ``Zakopane 1997, New developments in quantum field theory''
(hep-lat/9708023) ed. P. Damgaard et al. Plenum Press 1998;
C. Alexandrou, M. D'Elia and Ph. de Forcrand, Nucl. Phys. Proc. Suppl.
{\bf 83} (2000) 437 .
\end{thebibliography}
\end{document}